# [1]RCNN for Region of Interest Detection in Whole Slide Images

A Nugaliyadde[1], Kok Wai Wong[1], Jeremy Parry[1,2], Ferdous Sohel[1], Hamid Laga[1], Upeka V. Somaratne[1], Chris Yeomans[3], and Orchid Foster[3]

[1] Discipline of Information Technology, Mathematics and Statistics, Murdoch University, Perth, Australia.
[2] Western Diagnostic Pathology, Western Australia
[3] Pathwest Laboratory Medicine, Western Australia
A.Nugaliyadde@murdoch.edu.au

**Abstract.** Digital pathology has attracted significant attention in recent years. Analysis of Whole Slide Images (WSIs) is challenging because they are very large, i.e., of Giga-pixel resolution. Identifying Regions of Interest (ROIs) is the first step for pathologists to analyse further the regions of diagnostic interest for cancer detection and other anomalies. In this paper, we investigate the use of RCNN, which is a deep machine learning technique, for detecting such ROIs only using a small number of labelled WSIs for training. For experimentation, we used real WSIs from a public hospital pathology service in Western Australia. We used 60 WSIs for training the RCNN model and another 12 WSIs for testing. The model was further tested on a new set of unseen WSIs. The results show that RCNN can be effectively used for ROI detection from WSIs.

## 1   Introduction

Medical image processing is a complex task which requires complex image processing approaches [1] [2]. Medical images are larger and complex than many non-medical images used for image processing. However, often contain smaller variations in colour, hue and contours to an untrained human eye, making them challenging for computation because feature engineering is mostly based on general human perception of images [3]. Many image processing approaches have been applied for medical image processing for examples like X-Ray, magnetic, scopes, and thermal imaging [2] [4]. Whole Slide Image (WSI)s have been used extensively in digital pathology [5]. However, WSI presents unique challenges when compared to X-ray, CT scans and other medical images. WSIs' have high dimensions, show variation in stains between different WSIs, and often lacks label data, especially for ROI detection [6]. WSIs are very large, ranging from 3000 pixels x 4000 pixels to 55000 pixels x 60000 pixels, and the stains contrast between WSI can be substantial. Besides, WSIs often have a large area of background which are not of interest to pathologists [7]. Filtering out the background and the unwanted sections of WSI is an important step to assist pathologists in analysing the important regions of the image. The step can allow the pathologists to identify the Regions of Interest (ROI)s and perform more focused diagnosis using the identified ROIs [8]. The identification of ROIs is important and beneficial for further processing and analysis of images because it will act as a filter to pass only the ROIs to the pathologists, thus reducing the time spent on analyzing and processing of the images [9]. For example, Fig.1 illustrates that the identification of the ROIs can help the pathologists to identify the germinal center more accurately.

---

[1] This is a pre-print of the paper that was accepted by the 27th International Conference on Neural Information Processing 2020, and is a will be published in the CCIS Springer Series.

Segmentation on WSI to identify ROI is a common approach that has been developed over the last few years [10] [11] [12]. Most of the ROI detection methods are unsupervised because the number of training data is limited to WSI ROI detection [13] [14]. Segmentation requires the image to have similar variations throughout each WSI. However, in real-world WSIs, the staining is different and therefore, the segmentation parameters change from WSI to WSI [10]. Machine learning approaches require the use of feature engineering to facilitate identification tasks. However, most features in WSI are not easily visible or explainable by the experts. Therefore, a deep learning approach which learns features automatically show high potential [5] [15]. Convolutional Neural Network (CNN)s have a high potential in learning features without feature engineering in images [16] [5]. RCNN has emerged as a successful approach to learn and identify ROIs in images of many application domains [17] [18]. However, the variations in colour, hue and contour in most images are much higher and more apparent than WSI [19] [6]. Furthermore, most applications in other domains using RCNN to identify ROIs require the use of a large number of training data for learning [20] [21]. The purpose of this paper is to investigate the use of RCNN for ROI detection by using a small number of labelled WSIs for training.

The ROI selected for this study were Germinal Centres (GC) within normal and benign lymph nodes (Fig.1). GC are organized collections of activated lymphocytes and other immune cells that develop within follicles in response to immune stimulation. Before becoming stimulated, follicles lack GC and are called primary follicles. After stimulation, they develop GC and are called secondary follicles. Distinguishing primary and secondary follicles can be challenging for a pathologist, who may have to revert to using special antibody stains such as BCL6, which highlights key GC cells. This study aimed to develop and test an algorithm that can support a pathologist in identifying GC without using special stains. The real WSIs of patients used in this study is provided by a public hospital pathology service in Western Australia, and pathologists validate results.

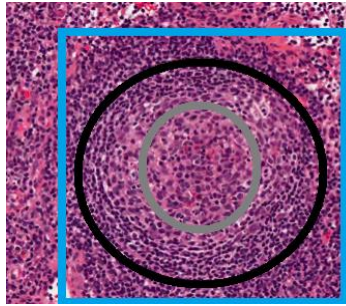

**Fig. 1.** The image shows the ROI that the pathologist marks. The blue box indicates the ROI, which encloses the GC boundary in black and GC in grey.

## 2    Methodology

The RCNN proposed by Girshick et al. was used for the experiment [17]. Fig 2 provided an illustration of the RCNN used for ROI detection in WSI. First, the large WSIs were patched, and the patches were passed through selective search to identify regions of proposals as described more in section 2.1. The candidate region proposals were moved onto a CNN. VGG16 pre-trained on ImageNet was used for the CNN because it is capable of capturing the basic features such as colour, hue, contours, etc. of any image. This feature extraction supports a model in learning features from a limited number of training data because the base of the feature extracted is already learnt, and only fine-tuning of the model is required. The CNN extracts the features (colour, hue and contours) of candidate regions, and the last layer is a dense layer which classifies the ROI and background. The proposed candidate regions and the ground truth calculate the intersection over union and label them. The model consists of two sections, which are sequentially connected to each other. The first section comprises of independent region proposal, which is used to extract the regions of an image. The second section is a large CNN, which extracts feature vectors and uses the feature vectors to classify the regions. The RCNN model in this paper learns to classify two classes; ROI and background. Although the general architecture and structure of the RCNN were used, adjustments were made to improve the ROI detection in WSIs.

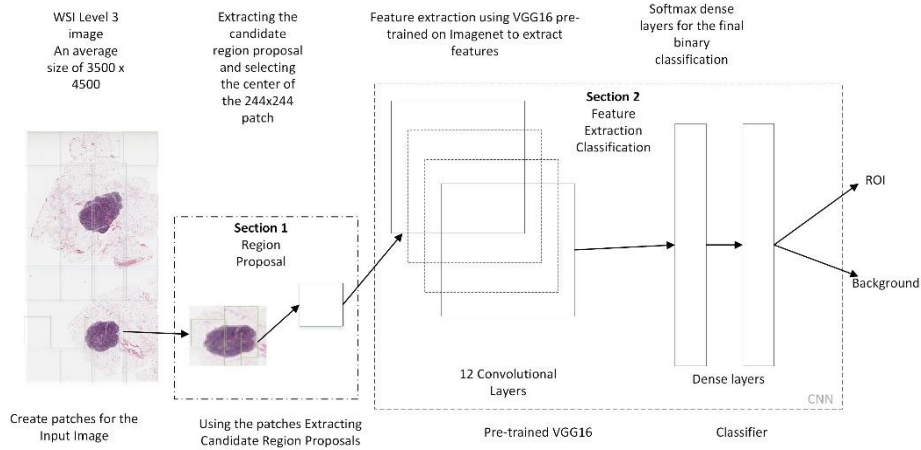

**Fig. 2.** Illustration of the flow of the RCNN method.

### 2.1 Region Proposal

Region proposal was used to avoid selecting many regions as potential ROIs for feature extraction. In the model, the region proposal generates candidate region proposal areas irrespective of any category. The patches are passed through a selective search, which is used to identify the candidate region proposals. The selective search will generate initial sub-segmentation for the initial candidate region proposals after which the similar regions are combined recursively to create larger candidate region proposals using the greedy algorithm. Finally, the generated regions are used to create the final candidate region proposals to be used for feature extraction and classification, which is the CNN, as shown in Fig.2.

### 2.2 Feature Extraction and Classification

Adjustment of the original RCNN in this paper focus on the feature extraction and classification layers. The candidate region proposals were passed to the CNN, and the CNN extracted 4096-dimensional features vector from each candidate proposal region. A 224 x 224 RGB image patch from the WSI was passed through 5 convolutional layers and two fully connected layers. The CNN used was based on the pre-trained VGG 16 model from ImageNet. The first 15 layers were frozen during the training process. The last-second layer was removed and replaced by a 2-unit softmax dense layer to predict the background and the ROIs. Adam optimizer was used with a learning rate of 0.0001. Categorical cross-entropy was used as the loss function. The final model had a total of 126,633,474 trainable parameters and 7,635,264 non-trainable parameters.

## 3    Experimentation and Results

In this paper, WSIs of patients from a public hospital pathology service in Western Australia are used in the study. We used 60 WSIs for training the RCNN model and another 12 WSIs for testing. The ROI that need to identify for this study were regions that could contain GC within normal and benign lymph nodes.

The first step of using the RCNN is to generate the appropriate patches. A sliding window which moved from left to right and top to bottom without any overlap was used to create patches of 244x244 pixels. These patches contained the marked ROIs by the pathologists. One of the objectives of this study was to find the best process of feeding the information into the designed RCNN model. Therefore, two experiments were set up. The first case, named as the Base Case, fed the entire patch (244x244) generated from the ROI into the RCNN to learn and predict. The second case, named as the Center Case, made use of the extracted version of the patch by taking 199x199 from the entire patch (244x244). The Center Case using 199x199 was selected after a trial had been performed to find the optimum centre patch size, in which199x199 provided the best performance. The following summarizes the two cases shown in this paper:

1. Base Case RCNN: The model used the entire patch (244x244) from the ROI to learn and predict.
2. Center Case RCNN: The model used the centre patch of 199x199 patch extracted from the centre patch of 244x244 to learn and train the model.

Table 1 shows a comparison between the two different approaches of RCNN models that are used for the ROI detection based on the test data set (12 WSIs). Intersection Over Union (IOU) is used to compare the model's results [22]. The Center Case RCNN outperformed the Base Case RCNN model for ROIs identification. The same training and testing data were used for both the models. The improved results demonstrate that considering the centre of the patch can support better ROI detection in WSIs used in the experiment. Fig. 3 presents the comparison of the ROIs identified by the Base Case RCNN and the Centre Case RCNN with the ground truth ROI of a testing WSI.

**Table 1.** IOU comparison between the different RCNN models used for ROI detection

| Base Case RCNN | Centre Case RCNN |
|---|---|
| 0.61 | 0.92 |

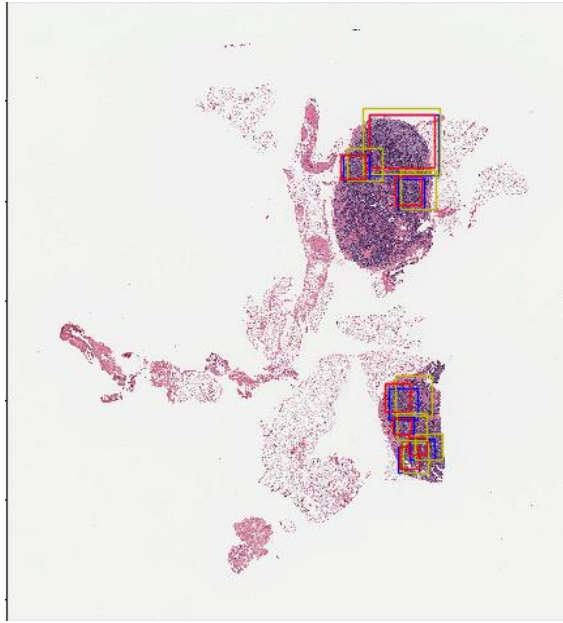

**Fig. 3.** The comparison of the ROIs generated from Base Case (yellow), Centre Case (blue) and the ground truth marked by pathologist (red) of a testing WSI.

## 4 Unseen Data Testing

After the RCNN model has been established and finalised from the previous experiment, the model was further tested on unseen data consist of 6 WSIs from the hospital. The 6 WSIs were given to a technical assistant and the trained RCNN model. The technical assistant and the RCNN model both annotated ROIs for the given WSIs independently. After which, a senior pathologist will evaluate and compare both annotations. The senior pathologist compared and evaluated the results, as shown in Table 2. The established RCNN identified a total of 112 from the 115 ROIs from the 6 WSIs, including some which were missed by the technical assistant. The discrepancy was estimated visually by directly comparing human ROI identification and RCNN ROI identification. Fig.4 shows a comparison of the RCNN identification with the human identified ROIs, and this shows both the model and human was able to locate all the ROIs.

From Table 2, it can be observed that specimen 2 and 6 were labelled as identical to those by the technical assistant, and validated visually by the senior pathologist. The results demonstrated that the RCNN model is capable of learning the features of the ROIs in WSIs from the 60 WSIs used in training and perform well for the testing set and the unseen dataset.

**Table 2.** Comparison of the results of the human technical assistant and the established RCNN.

| Specimen | Total ROI by Human | Total ROI by RCNN | Discrepant ROI |
|---|---|---|---|
| 1 | 44 | 44 | 4 |
| 2 | 8 | 8 | 0 |
| 3 | 12 | 12 | 4 |
| 4 | 14 | 14 | 1 |
| 5 | 17 | 14 | 8 |
| 6 | 20 | 20 | 0 |
| **Total** | **115** | **112** | **17** |

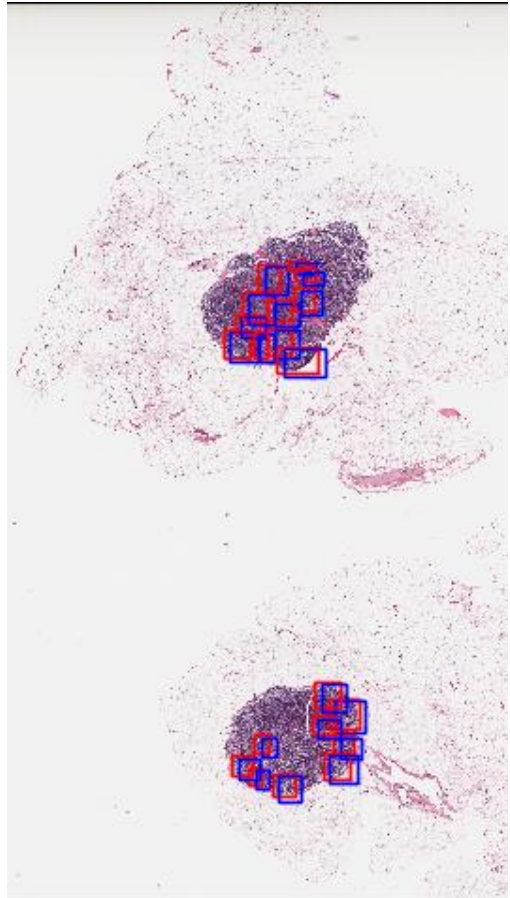

**Fig. 4.** This compares human annotations and the model annotations for Specimen 6. The model has predicted all the ROIs. Blue ROIs are marked by the RCNN model and red ROIs are marked by the technical assistant.

## 5     Discussions

Table 1 shows that the Centre Case performs better than the Base Case by obtaining an IOU of 0.92 (base case 0.61) for the RCNN that has been established in this paper. From observation, it was found that in the case of the Centre Case, the patch used to give a feature-rich area for the RCNN to learn ROI specific features. The use of the centre of a patch provided the model with a clearer ROI particular features. Therefore, the model was able to extract and learn the features of the ROIs accurately. In the unseen data test, the RCNN performed well as validated by the senior pathologist. The senior pathologist makes the decision by considering whether the ROIs identified have included the GC and its boundary (Fig.1). Therefore, even if the technical assistant's annotations and the RCNN's annotations were not 100% matching, the senior pathologist would consider that the ROIs have correctly been identified. In this case, the exact alignment of the ROIs identified by the technical assistant and the RCNN is not required.

In the unseen data testing, as evaluated by the senior pathologist, the established RCNN was capable of performing similarly to the technical assistant in Specimen 3 and Specimen 6, as shown in Fig.4 (Specimen 6) and Table 2. Furthermore, the proposed method was capable of identifying ROIs which were not identified by a human, technical assistant but missed ROIs that the technical assistant can identify (Fig.5). From Table 2, the differences are small as validated by the senior pathologist for other specimens.

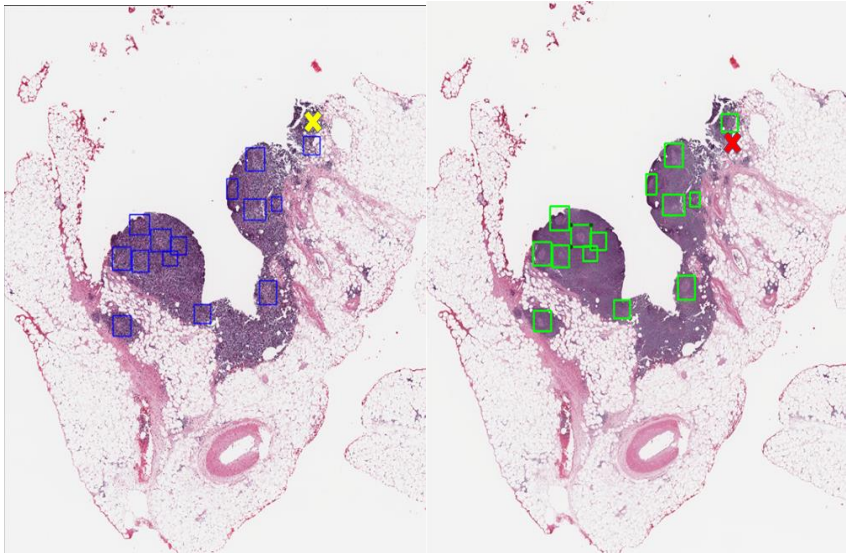

**Fig. 5.** The comparison between the technical assistant missing ROI and the RCNN model missing ROI in Specimen 4. The red cross indicates the ROI that the technical assistant missed, and the yellow cross indicates what the model missed.

## 6      Conclusion

In this paper, an investigation of the application for RCNN for WSI ROI identification is presented. RCNN's feature extraction and classification were modified for ROI detection in WSI using a limited number of training data. A public hospital pathology service in Western Australia provided the labelled WSIs. 60 WSIs and 12 WSIs were used to train and test the RCNN, respectively. Patches were made from the gigapixel images. The centres of the patches were used to train and test the RCNN. The use of the centre patch enabled the RCNN to learn features of the ROI well. Selective search, with the use of the greedy algorithm, was used to generate the candidate region proposal, and features were extracted using VGG 16 pre-trained on ImageNet, with the final softmax dense layer used to generate the classification. Results show that the established RCNN can be used to identify ROI on WSI, which could assist pathologists in the detection of regions that could contain GC within normal and benign lymph nodes.

Further work is underway to use a special protein maker stain to establish definitive ground truth for the germinal centre identification. This stain can be used for a comparison of human versus computer identification of the ROIs.

**Acknowledgements**. This project is funded by a Research Translation Grant 2018 by the Department of Health, Western Australia.